\documentclass[conference]{IEEEtran}
\IEEEoverridecommandlockouts
\usepackage{cite}
\usepackage{amsmath,amssymb,amsfonts}
\usepackage{algorithmic}
\usepackage{graphicx}
\usepackage{textcomp}
\usepackage{xcolor}
\usepackage{makecell}

\def\BibTeX{{\rm B\kern-.05em{\sc i\kern-.025em b}\kern-.08em
    T\kern-.1667em\lower.7ex\hbox{E}\kern-.125emX}}

\begin{document}
\title{Automatic Transformation of Natural to Unified Modeling Language: A Systematic Review}

\author{\IEEEauthorblockN{Sharif Ahmed}
\IEEEauthorblockA{\textit{Dept.  of Computer Science } \\
\textit{Boise State University}\\
Boise, USA \\
sharifahmed@u.boisestate.edu}
\and
\IEEEauthorblockN{Arif Ahmed}
\IEEEauthorblockA{\textit{Dept.  of Computer Science} \\
\textit{Boise State University}\\
Boise, USA \\
arifahmed@u.boisestate.edu}
\and
\IEEEauthorblockN{Nasir U. Eisty}
\IEEEauthorblockA{\textit{Dept.  of Computer Science} \\
\textit{Boise State University}\\
Boise, USA \\
nasireisty@boisestate.edu}
}

\maketitle

\begin{abstract}
\emph{Context:} 
Processing Software Requirement Specifications (SRS) manually takes a much longer time for requirement analysts in software engineering. Researchers have been working on making an automatic approach to ease this task. Most of the existing approaches require some intervention from an analyst or are challenging to use. Some automatic and semi-automatic approaches were developed based on heuristic rules or machine learning algorithms. However, there are various constraints to the existing approaches to UML generation, such as restrictions on ambiguity, length or structure, anaphora, incompleteness, atomicity of input text, requirements of domain ontology, etc.
. 
\emph{Objective:} 
This study aims to better understand the effectiveness of existing systems and provide a conceptual framework with further improvement guidelines.  
\emph{Method:} 
We performed a systematic literature review (SLR). We conducted our study selection into two phases and selected 70 papers.
We conducted quantitative and qualitative analyses by manually extracting information, cross-checking, and validating our findings.
\emph{Result:} 
We described the existing approaches and revealed the issues observed in these works. We identified and clustered both the limitations and benefits of selected articles. 
\emph { Conclusion: } 
This research upholds the necessity of a common dataset and evaluation framework to extend the research consistently. It also describes the significance of natural language processing obstacles researchers face.  
In addition, it creates a path forward for future research.

\end{abstract}


\section{Introduction}
\label{sec:Introduction}

\begin{figure*}[t]
\centerline{\includegraphics[trim={3cm 5.5cm 0 0},clip, width=\linewidth]{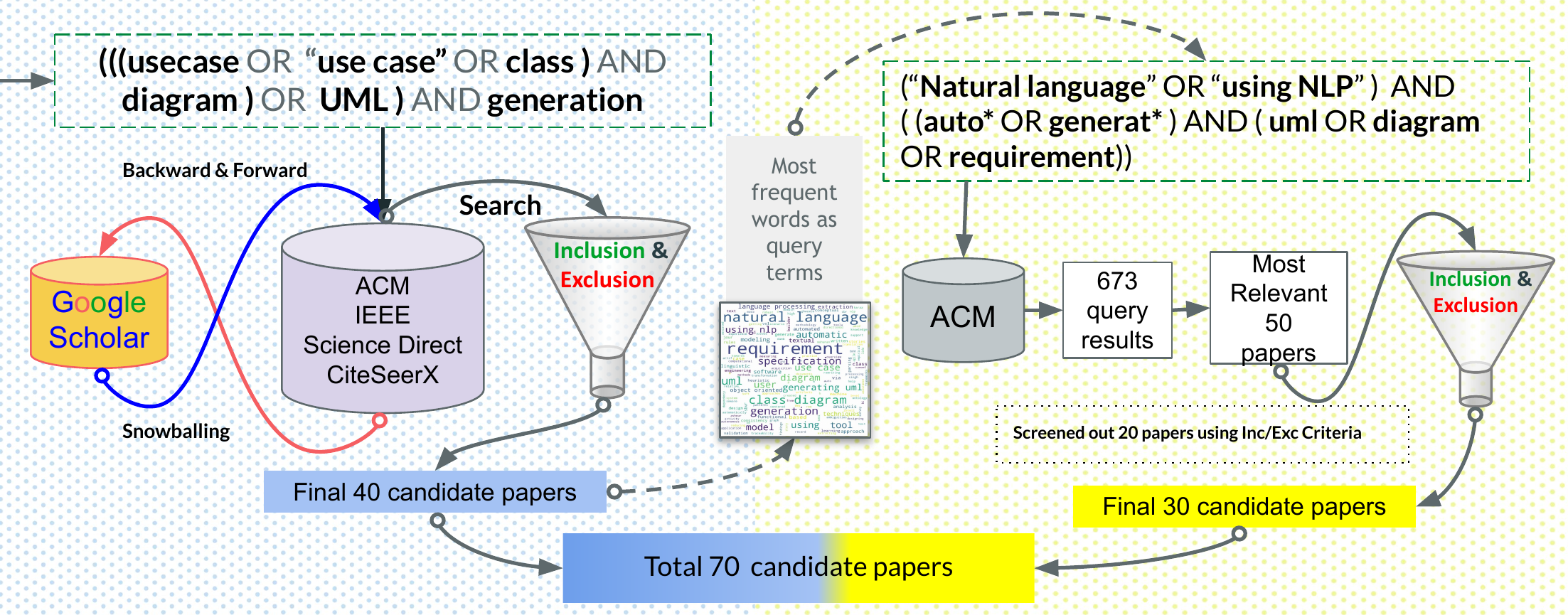}}
\vspace{-2mm}
\caption{Study selection process }
\label{fig_process_diagram}
\vspace{-4mm}
\end{figure*}

Requirements analysis in software development is an essential and rudimentary task. The quality analysis of requirements elicitation and further developing software work-products such as various diagrams and formatted textual descriptions leads to a successful project, product, or service. We human beings, as an analyst, have the innate power to understand the information in texts contextually regardless of the existence of misspelled words, incorrect grammar, and indirectly stated items. This performance has not been obtained yet in  Natural Language Processing (NLP) or Artificial intelligence, and it is a long way to go. But we human beings are also limited by fatigue, drowsiness, distractions, fluctuation in concentrations, which makes us error-prone despite our aforementioned power. So, the machines' ability to work relentlessly with error and humans' ability to work perfectly for short periods is not as easy to put on two pans of a weighing scale's beam and come to a conclusion. Berry et al. \cite{b1} suggested that NLP tools for requirements engineering can be an effective approach to activities that are tedious and rigorous.

Unified Modeling Language (UML) is ``a general-purpose visual modeling language to  specify, visualize, construct, and document the artifacts of a software system"- Rumbaugh et al. \cite{b88}. There is no fully working method or tool to generate  UML diagrams from the informal natural language (NL) requirements. Most existing approaches have high-level complexity, are computationally costly, or have limitations. Researchers found that earlier approaches require developer intervention for UML diagram creation. Besides, there are a few approaches that are comparatively recent and are fully automated. Still, these approaches are working under some constrained input such as restricted NL or some particular form of texts \cite{b4}. A previous work takes informal NL text requirements and generates UML Use Case and Activity Diagrams without any developer’s intervention or assistance \cite{b3}. Any fully automatic approach or tool can significantly help with requirement elicitation and modeling for software analysts. It can also save the overall cost and time within a software development process.

Many tools and approaches have been proposed and developed to extract information from natural language text to generate UML diagrams. However, natural language processing has many problems such as ambiguity, uncertainty, incompleteness, and incoherence  \cite{b4}. Existing approaches perform poorly at times because of this ambiguity of NL. On top of that, large requirement documents can produce other issues while processing. Besides, crucial information is sometimes missed, which feeds the verb to develop complete understanding. A software analyst has to figure out such problems and fix these issues using her domain knowledge. Many tools and techniques have been proposed to solve this kind of problem. But in reality, these tools are not being used in Software Development Life-Cycle for many constraints. And most of the tools require frequent interventions of an analyst to finish the process. In addition, the majority of tools are limited to producing class diagrams only \cite{b18, b19}.

Though there are some new solutions, they still require more rules to increase the domain knowledge, which is another problem \cite{b2, b4}. So, our work will provide a conceptual framework showing an overview of using rule-based heuristic approaches and machine learning-based approaches. This framework will help researchers to make decisions while finding a new solution.

\section{Related Work}
\label{sec:Background}
\begin{table*}
\centering
\caption{Protocol summary}
\label{tbl_protocol}
\begin{tabular}{|p{0.14\linewidth}|p{0.85\linewidth}|}
\hline
Research Questions & \makecell[l]{
 RQ1: What are the existing approaches to automate the UML generation?\\
  RQ2: 
  How effective are the existing approaches?}\\\hline

Search string &
  \makecell[l]{\textbf{Phase-1:} 
  (((usecase OR  ``use case" OR class ) AND diagram ) OR  UML ) AND generation\\
  \textbf{Phase-2:} 
  (``Natural language" OR ``using NLP" )   AND ( (auto* OR generat*) AND ( uml OR diagram OR requirement))} \\\hline

Search strategy &
  \makecell[l]{\textbf{Phase-1:} 
  DB search: ACM, Google Scholar, IEEE, ScienceDirect, Springer, CiteSeerX \\
  Backward and forward snowballing using Google Scholar\\
  \textbf{Phase-2:} 
  DB search: ACM\\
  Formulated new query string from most frequent words found in the titles of articles filtered in Phase-1} \\\hline

Inclusion criteria &\makecell[l]{ The article is written in English\\
  The article is published in a scholarly journal or conference/ workshop/ symposium proceedings\\
  The article covers at least one of our RQs partially or fully}\\\hline

  Exclusion criteria &\makecell[l]{
  The article doesn't answer any of our RQs either partially or fully\\
  The article is written in Non-English\\
  Any article retracted from publisher \\
  The article is secondary study derived from primary studies i.e. Systematic Literature Review or  Mapping Study} \\\hline

Study type &
  Primary studies 
\\\hline
\end{tabular}
\vspace{-4mm}
\end{table*}
Though there are not many systematic studies related to our research questions, we found a few notable literature reviews that we discuss in this section.

Zaho et al. \cite{b53} conducted a robust systematic literature review on NLP for requirements engineering (RE), capturing almost every aspect of RE tasks: detection, extraction, classification,  modeling, tracing and relating, search, and retrieval. They covered the breadth of RE, which is valuable for the software research community. In contrast, we give a depth of RE regarding NL to UML automation. As they covered on breadth, it is not expected to have a depth of any of the tasks mentioned above they covered.

Omer et al. \cite{b87} performed a survey that addressed techniques and outputs. Their analysis of strengths and weaknesses is limited to the number of diagrams each study generated. 

Esra et al. \cite{b4} conducted another systematic mapping study on the techniques and approaches of NL to UML Class diagram. However, our analysis is neither limited to class diagrams nor mapping the techniques and approaches. Instead, we dug deeper into the existing approaches, tools, pros, and cons. In addition, we cover what the primary studies addressed,  attempted, or overlooked and their success and failure anecdotes from their published works.

\section{Research Methodology}
\label{sec:Methodology}
In this section, we discuss our research methodology for this study. 
At first, we developed a research protocol following guidelines for systematic literature reviews by Kitchenham~\cite{s1} to make our research process uncompromising and evident. Then, we followed the following steps: research question formation, study selection, data extraction, and data synthesis. Table \ref{tbl_protocol} shows the protocol overview that we developed and executed.

\subsection{Research Questions}
Our research motivation and objectives led us to the following research questions:

\textbf{RQ1: What are the existing approaches to automate the UML generation?}  The answer to this question will recapitulate the existing tools and techniques used by the software engineering community. 

\textbf{RQ2: How effective are the existing approaches?}
This inquiry will portray opportunities and obstacles of ongoing effort to solve NL to UML transformations.

\subsection{Searching Strategy}
We mainly focused following digital libraries for our study:
\begin{itemize}
\item ACM Digital Library (ACM)
\item IEEE Xplore Digital Library(Xplore)
\item Science Direct Digital Library (Elsevier)
\item CiteSeerX
\item Google Scholar
\end{itemize}
Our study comprises two phases. At first, we searched on aforementioned digital libraries using the query string, \emph{(((usecase OR  ``use case" OR class ) AND diagram ) OR  UML ) AND generation}, addressing the most commonly used UML diagrams. We snowballed backward and forward with the retrieved candidate papers on Google Scholar. At that point, we selected 40 articles perusing the full articles. Then we checked a shallow synthesis of words from the titles and the abstracts of these papers and identified the top frequent words. In the second phase, using these words, we built another search string \emph{(``Natural language" OR ``using NLP") AND ( (auto* OR generat* ) AND ( uml OR diagram OR requirement))} and performed the second phase of our search on abstracts only. We chose only ACM Digital Library in this phase as it retrieved 673 articles, the least number of articles than other libraries. We manually selected 30 articles after screening the title and abstract from these retrieved articles. Our searching period was Sep 2021 to Dec 2021, and the publishing years of papers were from 1994 to 2021.

\subsection{Quality Assessment and Inclusion/Exclusion}
We considered the studies written in English; published in a scholarly SE journal, conference, workshop, or symposium proceedings. We checked if any of the studies already answered any of our RQs partially or fully for assessing quality. We excluded secondary studies, i.e., Systematic Literature Review and Systematic Mapping Study. We also deduplicated the papers as depicted in Fig.~\ref{fig_process_diagram}.

\subsection{Data Extraction}
As our final primary studies were selected, we developed a data extraction form to answer our RQs. Table \ref{tbl_data_extraction_form} shows data fields and their mapping to RQs. The first two authors manually went through each of the papers and extracted information from each of the primary studies to fillup the form.
The third author reviewed the results. 
Then we discussed the emerged conflicts and fixed them. 
\begin{table}
\centering
\caption{Data Extraction Form}
\label{tbl_data_extraction_form}
\begin{tabular}{|p{0.19\linewidth}|p{0.51\linewidth}|p{0.18\linewidth}|}

\hline
\textbf{Field}      & \textbf{Categories}                                          & \textbf{Relvant RQ} \\ \hline
  
  
Paper Title         & Free text                                                    & -                   \\ \hline
  
 Year    & Number                                                       &Demographics\\ \hline
  
Source              & Venue / Journal / Conference                                 &Demographics\\ \hline
  
Authors             & Free text                                                    & Demographics        \\ \hline
  
Abstract            & Free text                                                    & -                   \\ \hline
  
Relationships & (Aggregation, Inheritance, Generalization, Association, Composition, Dependency, Multiplicity, Inclusion, Exclusion) & RQ2 \\ \hline
  
Methodology 
& Analysis of their implementation                             & RQ2            \\ \hline
  
Pros                & Analysis of their resolved problem                           & RQ2            \\ \hline
  
Cons                & Analysis of their unresolved problem                         & RQ2            \\ \hline
  
Automation           & Automatic, Semi-Automatic, Manual                            & RQ1                 \\ \hline
  
Approach            & Heuristic Rule based, Machine Learning based, Hybrid (ML+HR) & RQ1                 \\ \hline
  
Output              & Type of  output(UML)produced
& RQ1 , RQ2           \\ \hline
  
Technology          & Analysis of the tools, technology used to solve the problem   & RQ2            \\ \hline
Evaluation & Free text & RQ1, RQ2 \\ \hline
\end{tabular}
\vspace{-6mm}
\end{table}

\subsection{Data Synthesis}
The data extraction phase led us to synthesize quantitative and qualitative results. Having extracted the data in the form, we synthesized it to discover: 
\begin{itemize}
    \item Year-wise  distribution of published studies
    \item Statistics of approaches to convert NL to UML 
    \item Statistics of automation to convert NL to UML 
    \item Relationships among UML components resolved 
    \item Distribution of  final/intermediate outputs acquired
    \item Usage of technologies or tools to solve the problems
    \item Analysis of dataset, metric, and evaluation techniques
    \item Analysis of strength and limitation of existing solutions
    
\end{itemize}

\begin{figure*}[!t]
    \centering
    \begin{minipage}{0.39\linewidth}
        \centering
        \includegraphics[trim={0  0 0 7.5cm},clip, width=2.7in]{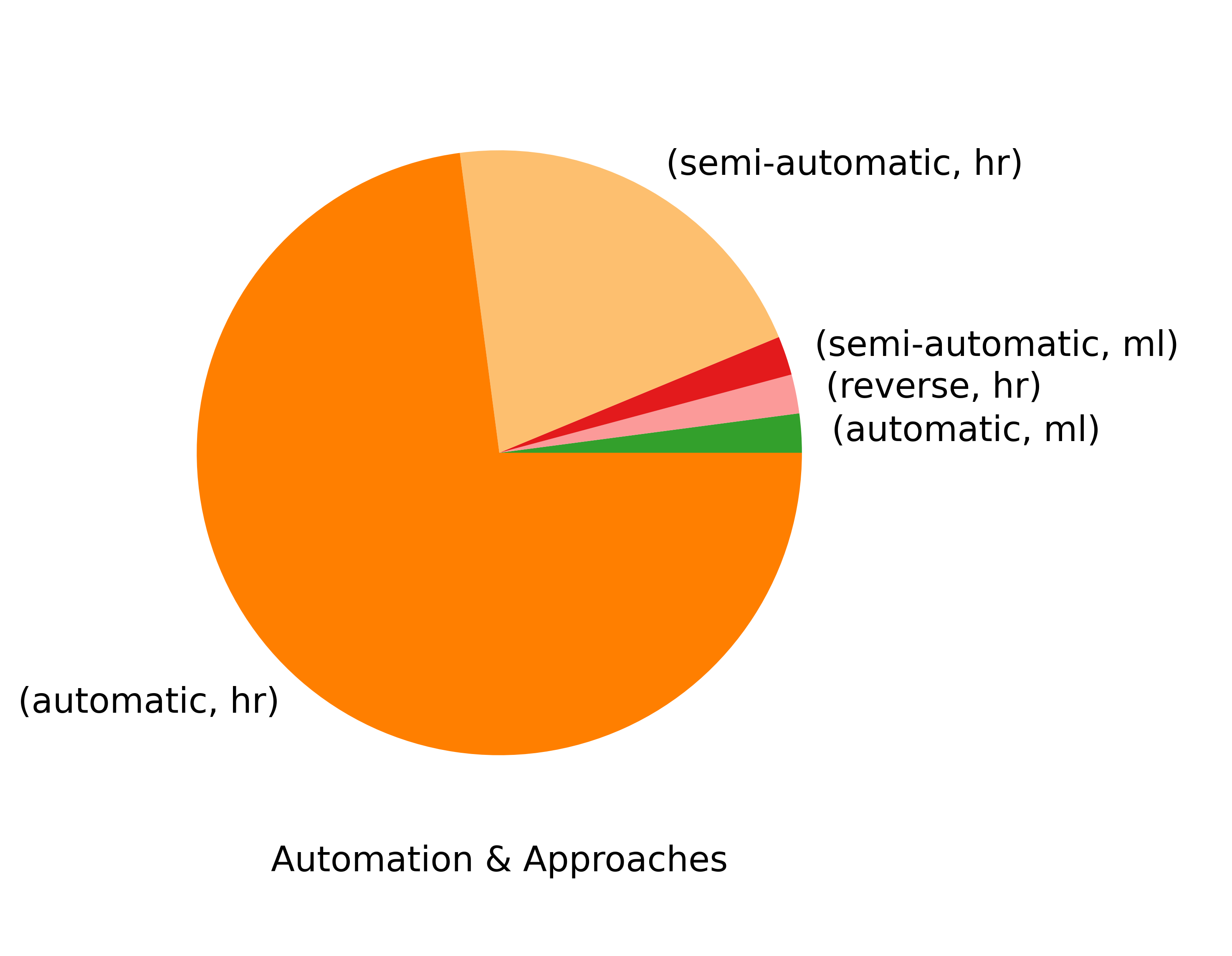}
    \end{minipage}
    \begin{minipage}{0.29\linewidth}
        \centering
        \includegraphics[trim={0  0 0 7.5cm},clip,width=2in]{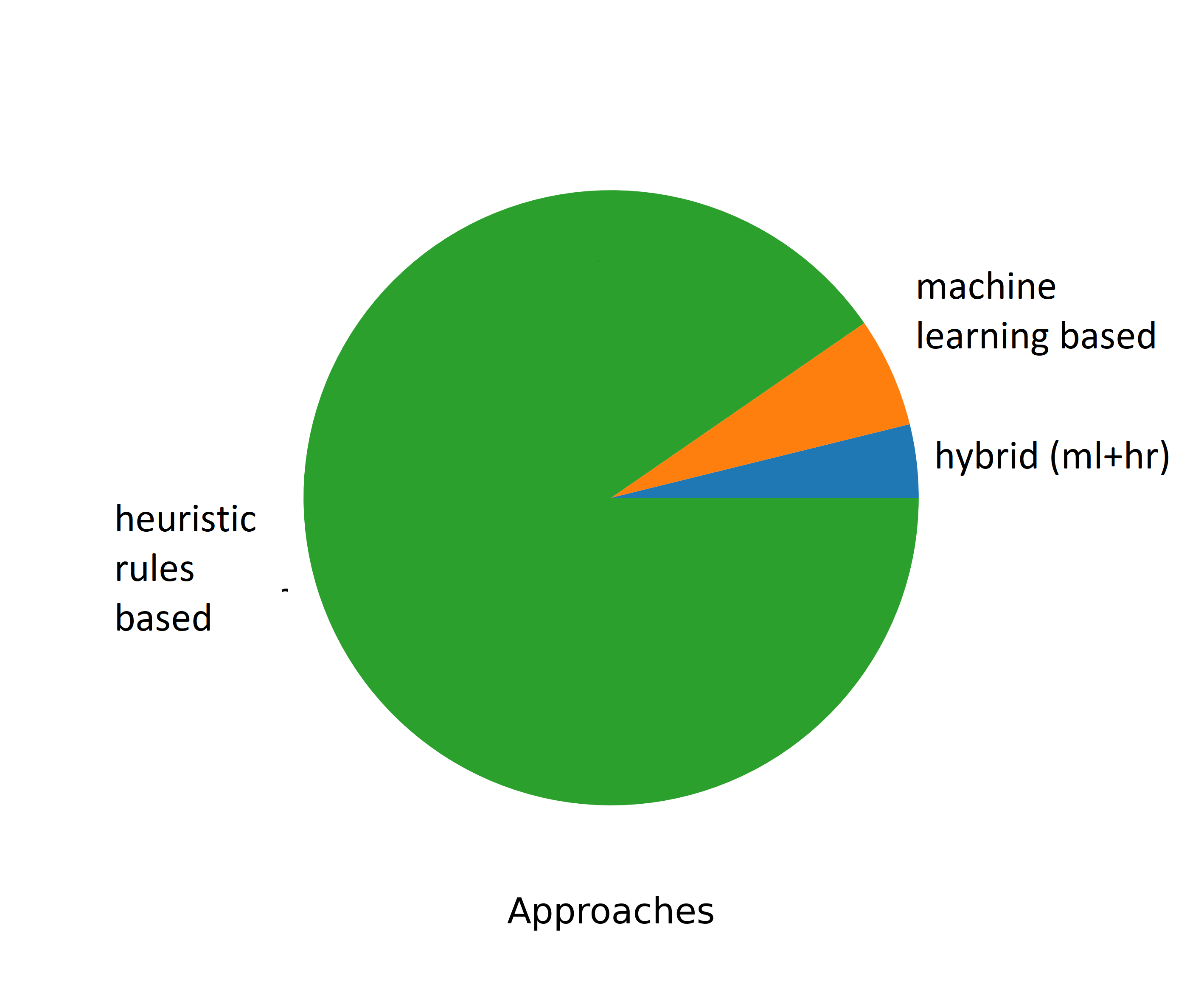}
    \end{minipage}
    \begin{minipage}{0.29\linewidth}
        \centering
        \includegraphics[trim={0  0 0 7.5cm},clip,width=2in]{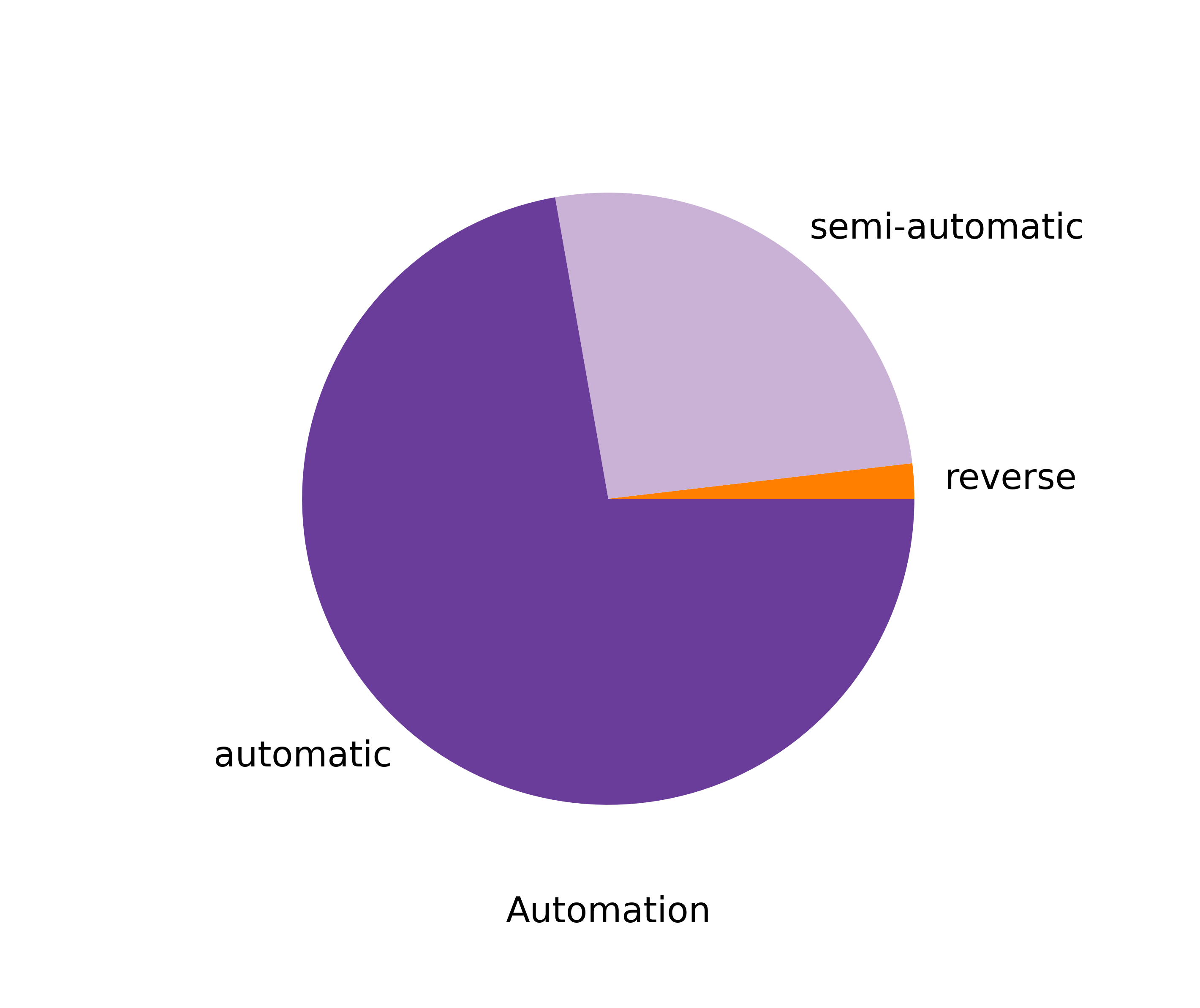}
    \end{minipage}
    \vspace{-2mm}
    \caption{Approaches to solve NL to UML transformation}
    \label{fig_approaches_pie}
    \medskip
\small
Here, \textbf{hr}: Heuristic rules based approach, \textbf{ml}: Machine learning approach
\vspace{-4mm}
\end{figure*}

The  Tables \ref{tbl_outputs}, \ref{tbl_technology}, \ref{tbl_relation_resolved}, and \ref{tbl_metrics} and Fig~
\ref{fig_approaches_pie} 
provide quantitative insights. 
Section 4.2 provides 16 facets with both quantitative and qualitative insights.

\section{Results}
\label{sec:Results}
We discuss our findings according to the research questions in this section. 

  \subsection{RQ1: What are the existing approaches to automate the UML generation?}
 
At first, we grouped all the approaches found in our primary selected papers into the following categories: Heuristic Rule-Based, Machine Learning, Automatic, Semi-automatic, and Hybrid.
Fig.~\ref{fig_approaches_pie} shows the approaches and automation we found in our analysis.
But as we moved forward, we figured out that many research works combined these approaches. So, we refined our cluster represented in Fig.~\ref{fig_approaches_pie}. We found that most of the work was carried out by applying the heuristic rules, and a few other works used machine learning techniques. In addition, some of the papers obtained significant output by using both heuristic rules and machine learning techniques. 

\begin{table}

\caption{Outputs obtained in studies}
\label{tbl_outputs}
\begin{tabular}{|p{0.30\linewidth}|p{0.57\linewidth}|}

\hline
\textbf{Outputs}               & \textbf{Studies}\\
\hline
class diagram         &   \cite{b3,b6,b12,b13,b16,b17,b18,b19,b20,b21,b25,b26,b27,b28,b29,b30,b32,b33,b34,b35,b37,b39,b40,b41,b42,b43,b63,b64,b76, b79, b80, b81} \\ \hline
usecase diagram       &                                                                                                                                                                                                                                                                           \cite{b2,b8,b10,b13,b15,b36,b45,b68} \\ \hline
object diagram        &                                                                                                                                                                                                                                                                                                             \cite{b22,b23,b24,b40,b47} \\ \hline
processed sentences   &                                                                                                                                                                                                                                                                                                                         \cite{b55,b56,b60,b70} \\ \hline
sequence diagram      &                                                                                                                                                                                                                                                                                                                          \cite{b6,b16,b27,b68} \\ \hline
comparison            &                                                                                                                                                                                                                                                                                                                          \cite{b5,b53,b66,b67} \\ \hline
code                  &                                                                                                                                                                                                                                                                                                                                      \cite{b6,b21,b64} \\ \hline
use case              &                                                                                                                                                                                                                                                                                                                                                              \cite{b7, b79} \\ \hline
test case             &                                                                                                                                                                                                                                                                                                                                                  \cite{b7,b64} \\ \hline
processed srs         &                                                                                                                                                                                                                                                                                                                                                 \cite{b54,b59} \\ \hline
activity diagram      &                                                                                                                                                                                                                                                                                                                                                  \cite{b2,b74} \\ \hline
meta model & \cite{b7, b51} \\ \hline   
\multicolumn{2}{|p{0.95\linewidth}|}{
\textbf{Other Outputs: }
traceability\cite{b45}; graph\cite{b72}; 
sbvr\cite{b57}; 
er-diagram\cite{b11}; 
processed named entity\cite{b38}; 
b-spec\cite{b71}; 
owl class\cite{b48}; 
proposal\cite{b73}; 
collaboration diagram\cite{b13}; 
test cases\cite{b7}; 
natural language\cite{b52}; 
feature diagram\cite{b69}; 
gui prototype\cite{b58}
}     \\\hline
\end{tabular}
\vspace{-4mm}
\end{table}
\subsubsection{UMLs Generated} Table \ref{tbl_outputs} shows the UML diagrams that existing studies covered. 
Table~\ref{tbl_outputs}presents the most common UML outputs such as usecase, sequence, and activity. Among them, the class diagram is studied the most. 
\begin{table}
\centering
\caption{Technology used in existing studies}
\label{tbl_technology}
\begin{tabular}{|p{0.21\linewidth}|p{0.70\linewidth}|}
\hline
\textbf{Technology} &
  \textbf{Studies} \\ \hline
nlp & \cite{b6,b7,b8,b12,b15,b16,b17,b18,b21,b22,b23,b25,b26,b33,b34,b35,b36,b37,b38,b39,b40,b41,b43,b45,b47,b53,b54,b56,b57,b58,b60,b67,b68,b70,b72,b73,b74,b76}\\ \hline
rule &\cite{b1,b17,b21,b28,b30,b32,b39,b40,b41,b43,b45,b47} \\ \hline
pos tagger &\cite{b1,b8,b10,b11,b15,b17,b18,b19,b26,b33,b34,b35,b37,b70} \\ \hline
parse &\cite{b1,b5,b10,b11,b13,b15,b22,b28,b37,b68,b73} \\ \hline
stanford  corenlp &\cite{b1,b2,b5,b6,b13,b16,b17,b21,b36,b68,b72,b73} \\ \hline
 ontology &\cite{b1,b12,b15,b18,b25,b35,b37,b38,b39} \\ \hline
parser &\cite{b5,b13,b15,b22,b28,b37,b68,b73} \\ \hline
wordnet &\cite{b5,b6,b12,b13,b18,b35,b37,b59} \\ \hline
tree &\cite{b5,b10,b11,b15,b52,b63} \\ \hline
gui &\cite{b18,b23,b25,b33,b59,b60} \\ \hline
open  nlp & \cite{b12,b33,b35,b37,b73} \\ \hline
dependency &\cite{b20,b29,b63,b68} \\ \hline
sbvr &\cite{b17,b32,b17,b32} \\ \hline
graph &\cite{b29,b63,b72} \\ \hline
ml &\cite{b27,b69,b76} \\ \hline
ocl &\cite{b7,b27,b63} \\ \hline
traceability &\cite{b33,b34} \\ \hline

\multicolumn{2}{|p{0.95\linewidth}|}{
\textbf{Other Technologies: }
brill\cite{b18}; 
treetagger\cite{b15}; 
bayes\cite{b8}; 
featureide\cite{b69}; 
gkps\cite{b20}; 
javarap\cite{b13}; 
reasoning\cite{b26}; 
sharpnlp\cite{b60}; 
spacy\cite{b70}; 
spider\cite{b18}; 
verbnet\cite{b59}; 
semnet\cite{b80}; 
lolita\cite{b80}; 
rtool\cite{b79}; }\\\hline

\end{tabular}
\vspace{-6mm}
\end{table}

\subsubsection{Technologies Used} We have manually extracted the technologies used from each of the papers from our primary studies. Table \ref{tbl_technology} shows the list of technologies used. Researchers used a variety of NLP techniques in their studies. We observed significant use of the Stanford CoreNLP library. Besides, we also found the use of OpenNLP and SharpNLP. Stanford's POS tagger, Brill, TreeTagger, dependency parser, POS parser, and Stanford's parser were most commonly used for Parts of Speech (POS) tagging. WordNet was most common for resolving ambiguity, but WSD and VerbNet were also used.   
A set of heuristic rules were defined for heuristic-rule-based solutions. We also found several works using the domain ontology technique.

\subsubsection{Preprocessing techniques applied} We found that all the preprocessing treatments for NL are  almost similar in nature. The first task in preprocessing is to split the text into sentences. Then the sentences are tokenized using either stemming or lemmatization. Surprisingly we found several works that applied both techniques together \cite{b2,b3,b4}.
In stemming, the output can be meaningless words, affecting the use of tagger and parser. Then POS tagger and POS parser are used for identifying NP, VB, grammatical structures, etc. Then a set of rules is applied to identify actors, use cases, relationships, etc. But while normalizing the NL text automatically, there is a high chance of losing information \cite{b13}. So, the decision and treatment selection in preprocessing phase affects the overall performance.

\subsection{ RQ2: How effective are the existing approaches?}

\begin{figure*}[t]
\centerline{\includegraphics[clip, width=0.9\linewidth]{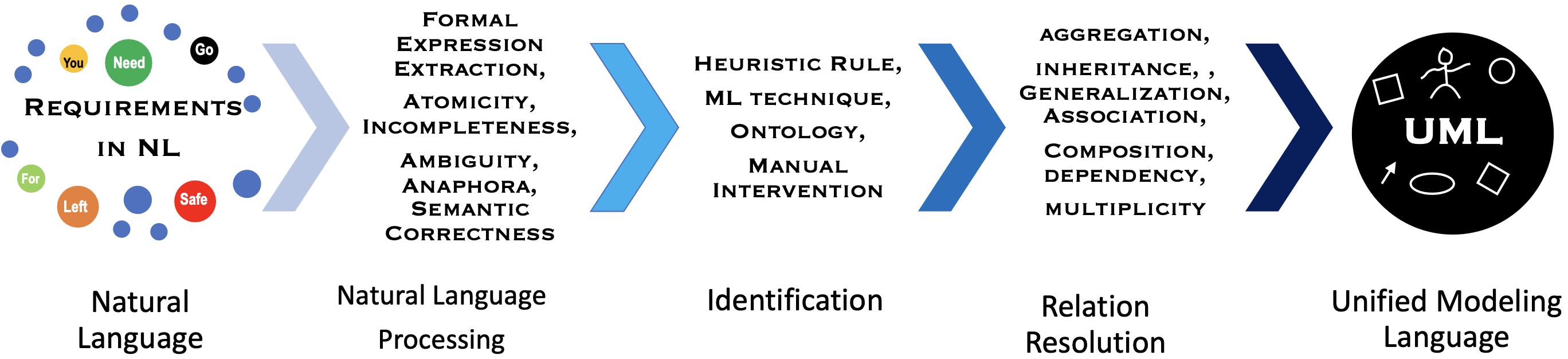}}
\vspace{-2mm}
\caption{Conceptual Framework }
\label{fig_framework}
\vspace{-4mm}
\end{figure*}

In this study, one of our objectives was to find ``to what extent problem(s) were solved" and ``what limitations were observed" from our primary studies. To do so, we extracted both pros and cons from each study. Then, we framed our findings into a conceptual framework comprising sixteen different facets in Fig.~\ref{fig_framework}. Finally, we describe how far researchers could solve existing problems and what problems they could not solve while translating NL to UML in this section. 

\subsubsection{Ambiguity} The requirement document contains text where comprehension or meaning extracted by readers may vary based on readers' perspectives. In particular, requirement analysts can resolve the ambiguity from text with their domain knowledge. But when it comes to NLP,  it becomes much harder to capture the correct meaning of a word by resolving ambiguity. Any misinterpretation of a word may produce defects that can affect the overall context \cite{b50}. Some studies solved ambiguities but mostly resolved lexical ambiguity. In NLP, syntactic or semantic ambiguities are also common problems. But, existing approaches could not resolve such ambiguities completely \cite{ b79,b80,b47, b54, b59,b66, b67}. 
 
A study stated six kinds of ambiguity in software requirements: lexical, syntactic, semantic, pragmatic, vagueness, and language error \cite{b86}. However, in their work, they only focused on lexical, syntactic, and language errors only \cite{b59}. We found some manual glossaries, machine learning, and ontology-based approaches to reduce ambiguity from the SRS. In addition, we found some tools for identifying and eliminating ambiguities either fully or partially. Tools are- WSD, QuaARS, ARM, RESI, SREE, NAI, SR-Elicitor, and NL2OCL \cite{b62}. 

\subsubsection{Semantic Correctness} We found several studies resolved semantic incorrectness by using WordNet. But, in a study, they considered frequency counts while using WordNet caused misclassification \cite{b12,b37,b45}.

\subsubsection{Language} We did not find any study processing Non-English requirement text. We only found a work where the requirement text was translated from French into English \cite{b5}. 

\subsubsection{Heuristic Rule} Studies using heuristic rules have shown promising work progress. However, most of them addressed the necessity of more heuristic rules for better performance \cite{b2,b3}. 

\begin{table}
\centering
\caption{Relationships resolved in existing studies}
\label{tbl_relation_resolved}
\begin{tabular}{|p{0.27\linewidth}|p{0.65\linewidth}|}

\hline
Relation Resolution & Studies \\
\hline
aggregation    &   \cite{b3,b4,b11,b12,b15,b16,b17,b18,b19,b20,b22,b32,b36,b37,b39,b41,b49,b50, b79} \\ \hline
association    &                           \cite{b3,b6,b12,b31,b19,b21,b22,b30,b32,b36,b38,b39,b40,b41,b49,b50,b69, b79, b80} \\ \hline
generalization &                                                                                                   \cite{b3,b4,b12,b17,b18,b19,b30,b32,b37,b39,b41, b79} \\ \hline
composition    &                                                                                                                                      \cite{b3,b11,b12,b31,b15,b32,b36,b37} \\ \hline
multiplicity   &                                                                                                                                                                         \cite{b15,b19,b32,b39,b40} \\ \hline
dependency     &                                                                                                                                                                                       \cite{b3,b4,b32,b37} \\ \hline
inheritance    &                                                                                                                                                                                                 \cite{b32,b36} \\ \hline
\end{tabular}
\vspace{-5mm}
\end{table}


\subsubsection{Relation Resolution}
Several relationships connect the UML components. For instance, the association relationship connects each actor and the use case. Meryem et al. \cite{b15} identified the actors first and then discovered the use cases using the linkage from the actors. And for more than one property associated with a concept or candidate, it is considered as a class; otherwise, an attribute \cite{b18}. Prepositions are emphasized for identifying relationships and associations \cite{b21}. Verb phrases are used to determine inheritance, association, composition, and aggregation relations to extract actors or classes; association and composition relations are used for identifying actions, operations, or methods \cite{b36}. Another type of relationship is semantic relation comprising vertical and horizontal relationships. Vertical relations include broader, part-of, or instance-of. Horizontal relations are similarity or relatedness \cite{b18}.
SBVR, which has Object Oriented information, easily retrieves the association, multiplicity, aggregations, generalizations, and instances. A few others used Named-Entity-Recognition, Stanford Open Information Extraction, Ontology, WordNet, and Dependency Parser to solve these relations.

\subsubsection{Text Restriction} We found a few papers that mentioned different constraints on NL text \cite{b8, b60}. We also investigated the structure of the NL text used for transforming into UML. Some studies put restrictions on:
\begin{itemize}
\item Text length: No study explicitly mentioned the text length except Sandeep et al.  \cite{b8}. On average there were 10 sentences per story which are comparatively shorter.  In reality,  the original user stories are much more lengthy and lack clarity. 
\item Sentence length: Some approaches only used 5-15 sentences with an average sentence length of 10 words in their user requirement document. Some studies observed that long sentences having more than one N (Nouns) or VB (Verbs) can create disturbingly long use cases \cite{b8}. 
\item Grammatical structure: The constraint to use modal verbs \cite{b60} or active sentences \cite{ b45, b13, b46, b35} only in requirement text imposes additional restrictions.
\item Controlled language: A few approaches were carried out using controlled NL instead of NL \cite{b17, b57, b66}.
\end{itemize}

\subsubsection{Formal Expression Extraction} In formal expression extraction, an expression must be consistent with the formal representation of the system. Dependency graph construction based on grammatical structure and abstract syntax trees can show similarities. But, synonyms applied within a sentence cause major problem \cite{b59}.
For example, ``CPU" and ``Processor" can be used within the same sentence. In that case, Word Sense Disambiguation (WSD) is a good choice to resolve such issues. 

\subsubsection{Dataset}

Most of the primary studies applied their models to one or more NL software requirement text(s) (aka case-study) to demonstrate the models' performance. We found input NL case-study appeared more than twice are: Library \cite{ b11, b12, b17, b20, b32, b41, b76}, ATM   \cite{ b79, b31, b68, b74}, Elevator \cite{ b22, b68, b74}, Banking \cite{ b29, b76, b33, b34}, Home \cite{ b69, b70}, Arena \cite{ b68, b74}. However, some studies didn't even address their input case-study \cite{ b8, b16, b17, b19, b31, b71}. The number of input case-study used in the experiment ranges from 0 to 7.  

Meryem et al. \cite{b15} worked with NLP techniques along with OWL ontology and MAT, then simulated the performance, including accuracy, recall, precision on classifying actor, use case, and relationships on a dataset that is obsolete and inaccessible. There are a few papers that mentioned the dataset they used. But we did not find all of them available publicly. We also found a few studies using significantly small and poor datasets \cite{b17, b64}.

\subsubsection{Anaphora Resolution} This is one of the most common and challenging problems in NLP. A few studies solved the only basic or pronominal issues using JavaRAP by replacing proper nouns with their correct noun form \cite{b13, b47}. 
For example: ``John found the love of his life" where `his' refers to `John'.

\subsubsection{Incompleteness} Requirement text with correct grammatical structure may have an information gap if the document is written considering readers understandability. Any human reader having field expertise can comprehend such incompleteness. But, it is difficult to capture such information gaps using incorporated NLP techniques \cite{b38, b50}.  
 
\subsubsection{Atomicity} Another issue comes with a lack of atomicity which appears when the writer keeps the text unambiguous by providing all the information to make the text complete but addresses more than one thing. In NLP, if we can split a sentence into smaller sentences, then it doesn't go along with atomicity. A few studies solved this problem by splitting a sentence into smaller sentences. But, for complex sentences, it required human intervention \cite{b50}.  

 
\subsubsection{Misclassification} Researchers found a high level of misclassification for classes and attributes for using simple heuristic rules \cite{b5, b20}. Abdelkareem et al. \cite{b16} normalized the text before POS tagging then used Stanford NLP POS Tree. But, they found that the Stanford PoS tree misses Sub (Subject) by misclassifying VB (Verb)/NP (Noun Phrase). 
 
\subsubsection{Manual Intervention} We found studies producing outcomes with and without manual intervention. A few studies require it mandatorily, some of them require it to some extent \cite{b1, b2, b6, b40}. 

\subsubsection{SRS Traceability} After identifying class, attributes, and operations (UML components), identified UML components are checked in a backward fashion to see whether the UML components are meaningful and working or not. This iterative process provides meaningful class diagrams \cite{b33,b34,b35}. 
 
\subsubsection{Proof of Concepts} Many of the studies provided various software, IDE plugins, or other forms of deliverables, which are essential for the research community to believe their demonstrations and feasibility \cite{b16,  b21,  b23, b12,  b57, b58}. 

\subsubsection{Evaluation}
\begin{table}
\centering
\caption{Metrics used for evaluation in existing studies}
\label{tbl_metrics}
\begin{tabular}{|p{0.36\linewidth}|p{0.57\linewidth}|}

\hline
Metrics & Studies \\ \hline
precision                     &                                                                                                                                                                                                                                                \cite{ b5, b8, b80, b15, b17, b19, b20, b31, b32, b33, b37, b39, b41, b57, b69, b70, b76, b83, b84} \\ \hline
recall                        &                                                                                                                                                                                                                                                             \cite{ b8, b80, b17, b19, b20, b31, b32, b33, b37, b39, b41, b57, b69, b70, b76, b83, b84} \\ \hline
enumeration                   &                                                                                                                                                                                                                                                                                                                                                 \cite{ b7, b29, b31, b33, b42} \\ \hline
accuracy                      &                                                                                                                                                                                                                                                                                                                                                 \cite{ b8, b10, b16, b59, b68} \\ \hline
relationship                           &                                                                                                                                                                                                                                                                                                                                                        \cite{ b3, b19, b39, b69} \\ \hline
false positive                            &                                                                                                                                                                                                                                                                                                                                                              \cite{ b70, b76, b84} \\ \hline
over specification                            &                                                                                                                                                                                                                                                                                                                                                              \cite{ b19, b20, b41} \\ \hline
output type                          &                                                                                                                                                                                                                                                                                                                                                                \cite{ b2, b3, b13} \\ \hline
false negative                            &                                                                                                                                                                                                                                                                                                                                                              \cite{ b70, b76, b84} \\ \hline
completeness                  &                                                                                                                                                                                                                                                                                                                                                                     \cite{ b54, b74} \\ \hline
f-measure                     &                                                                                                                                                                                                                                                                                                                                                                     \cite{ b57, b76} \\ \hline
                                                                                                    
\multicolumn{2}{|p{0.95\linewidth}|}{
\textbf{Other Metrics:} 
weighted average \cite{ b57}, f-measure\cite{ b57}, 
ambiguity\cite{ b54, b59}, 
over-generalization\cite{ b39}, 
consistency \cite{ b74},  
automation \cite{ b13}, 
atomicity\cite{ b54}, 
conceptual density of dataset \cite{ b83}, 
limitation                    \cite{ b79}, 
verifiability                 \cite{ b59}, 
computational time \cite{b54}
}
\\ \hline
\end{tabular}
\vspace{-6mm}
\end{table}
The most common metrics found in our primary studies are based on counting proper classification or extraction of UML components from NL such as Precision, Recall
Accuracy, F-measure, Enumeration,  Over-specification, Over-generalization False Negative, False Positive. A few other less common metrics found are Completeness,  Ambiguity,  Conceptual Density of Dataset, Computational time,  Weighted Average (in addition to enumeration), Verifiability,  Consistency,  and Atomicity. Some studies just contrasted their implementation with others as an evaluation such as Relationship, Output Type, Automation. 

\section{Discussion}
\label{sec:Discussion}


We have found most papers using a rule-based heuristic approach which indicates the common trend for solving, but machine learning approaches are also noticeable in recent years. Each of the studies had taken an almost similar treatment for NL preprocessing using different libraries. 


We observed many patterns in the existing research works and discovered trends and relationships. As described in section \ref{sec:Results}, we identified sixteen facets of the solutions and limitations of the existing approaches. We found the class diagram as the most used UML diagram. Implementing other UMLs like use case or usecase diagrams does require additional tasks and effort. We also found a comparatively low number of research mentioning their proof of concepts, but in some cases, these were neither publicly accessible nor found on the web. The primary concern for most of the studies was the ambiguity of NL. Few works have significantly mentioned this problem, its limitations, and its treatment. 

However, most implementations had constraints such as satisfying a specific grammatical structure, using a domain ontology, sentence length,  and absence of ambiguity or anaphora. Some also failed to identify UML components or detect relationships among them in some cases. Many evaluation metrics and datasets emerged in this study. Claims from evaluations are not comparable in many cases because of differences in metrics or datasets. Moreover, the case studies used lack more extensive or diverse descriptions.  

\section{Threats to Validity}
\label{sec:Threats}
We did not use SpringerLink and Scopus due to not having access to these digital libraries within the period of this study. The SLR was independently executed by the first two authors and reviewed by the third author to minimize personal bias. We might have some selection bias in phase-1. We analyzed the most frequent words from the titles of selected 40 papers from phase-1. Using these most frequent words as query terms, we again searched the same digital libraries. Thus, the later searching phase minimizes query term selection bias as we constructed it from statistics instead of our choice. Among them, ACM digital library retrieved the least number of papers. So we considered only ACM for the second phase and manually selected the most relevant 30 papers for our second phase study selection. The reduced number of relevant papers in the second phase indicates that our two-phased search strategy was effective and had good evidence coverage.  


\section{Conclusion}
\label{sec:Conclusion}
This systematic literature review framed research works that generated UML components from NL. We focused on approaches used, their pros and cons, and metrics. In addition, we identified the contributions and limitations of these works in this study. We found several heuristic-rule-based solutions. It can be a potential avenue to build a robust framework using machine learning techniques by exploiting these heuristic rules. 
In many research fields, several common standard datasets exist that help researchers extend their research work rapidly. We believe that establishing a benchmark dataset and designing metrics (quantitative and qualitative)  for evaluating NL to UML transformation will help the community.

This systematic literature review will help the researchers of the NLP and RE fields. This review pointed out major NL problems that researchers are trying to resolve. Furthermore, we have analyzed and conceptually framed the existing approaches and their pros and cons. Hence, our work will help both the researchers and developers working on the automation of NL to UML transformation.

\bibliographystyle{IEEEtran}
\bibliography{sigproc}
\end{document}